\documentstyle[aps,prd,multicol,epsfig]{revtex}

\begin{document}

\title{The Power Spectrum of the Sunyaev--Zel'dovich Effect}

\author{Alexandre Refregier\thanks{previously at the
Department of Astrophysical Sciences, Princeton University,
Princeton, NJ 08544}}
\address{Institute of Astronomy, Madingley Road, University of
Cambridge, Cambridge CB3 OHA, England; ar@ast.cam.ac.uk}

\author{Eiichiro Komatsu\thanks{also at the Astronomical Institute, 
Tohoku University, Aoba, Sendai 980-8578, Japan}}
\address{Department of Astrophysical Sciences, Princeton University,
Princeton, NJ 08544, USA; komatsu@astro.princeton.edu}

\author{David N. Spergel}
\address{Department of Astrophysical Sciences, Princeton University,
Princeton, NJ 08544, USA; dns@astro.princeton.edu}

\author{Ue-Li Pen}
\address{Canadian Institute of Theoretical Astrophysics, University of
Toronto, 60 St. George St., Toronto, Canada; pen@cita.utoronto.ca}

\date{To appear in PRD - submitted December 1999}
\maketitle

\begin{abstract}
The hot gas in the IGM produces anisotropies in the Cosmic Microwave
Background (CMB) through the thermal Sunyaev-Zel'dovich (SZ)
effect. The SZ effect is a powerful probe of large-scale structure in
the universe, and must be carefully subtracted from measurements of
the primary CMB anisotropies. We use moving-mesh hydrodynamical
simulations to study the 3-dimensional statistics of the gas, and
compute the mean comptonization parameter $y$ and the angular power
spectrum of the SZ fluctuations, for different cosmologies. We compare
these results with predictions using the Press-Schechter formalism.
We find that the two methods agree approximately, but differ in
details. We discuss this discrepancy, and show that resolution limits
the reliability of our results to the $200 \lesssim l \lesssim 2000$
range.  For cluster normalized CDM models, we find a mean
$y$-parameter of the order of $10^{-6}$, one order of magnitude below
the current observational limits from the COBE/FIRAS instrument.  For
these models, the SZ power spectrum is comparable to the primordial
power spectrum around $l=2000$. It is well below the projected noise
for the upcoming MAP satellite, and should thus not be a limitation
for this mission. It should be easily detectable with the future
Planck Surveyor mission. We show that groups and filaments ($k T
\lesssim 5$ keV) contribute about 50\% of the SZ power spectrum at
$l=500$. About half of the SZ power spectrum on these scales are
produced at redshifts $z \lesssim 0.1$, and can thus be detected and
removed using existing catalogs of galaxies and X-ray clusters.
\end{abstract}


\begin{multicols}{2}

\section{Introduction}

The hot gas in the IGM induces distortions in the spectrum of the
Cosmic Microwave Background (CMB) through inverse compton
scattering. This effect, known as the thermal Sunyaev-Zel'dovich (SZ)
effect \cite{sun72,sun80}, is a source of secondary anisotropies in
the temperature of the CMB (see Refs.\onlinecite{rep95,bir99,ref99a} for
reviews). Because the SZ effect is proportional to the integrated
pressure of the gas, it is a direct probe of the large scale structure
in the low redshift universe. Moreover, it must be carefully
subtracted from the primary CMB anisotropies, to allow the
high-precision determination of cosmological parameters with the new
generation of CMB experiments (see \onlinecite{zal97,bon97}
and references therein).

Thanks to impressive recent observational progress, the SZ effect from
clusters of galaxies is now well established
\cite{rep95,bir99,car99,gra99,kom99b}. The statistics of SZ clusters
were calculated by a number of authors using the Press-Schechter (PS)
formalism (eg. \onlinecite{mak93,bar94,col94,del95,bar96}). Recently,
Atrio-Barandera \& M\"{u}cket \onlinecite{atr99} used this formalism,
along with assumptions about cluster profiles to compute the angular
power spectrum of the SZ anisotropies for the Einstein--de Sitter
universe. A similar calculation was carried out by Komatsu \& Kitayama
(KK99, hereafter)\onlinecite{kom99}, who also studied the effect of
the spatial correlation of clusters and cosmological models.

The statistics of SZ anisotropies have also been studied using
hydrodynamical simulations. Scaramella et al. \cite{sca93}, and more
recently da Silva et al. \cite{das99}, have used this approach to
construct SZ maps and study their statistical properties.  Persi et
al. \cite{per95} instead used a semi-analytical method, consisting of
computing the SZ angular power spectrum by projecting the
3-dimensional power spectrum of the gas pressure on the sky.

In this paper, we follow the approach of Persi et al. using Moving
Mesh Hydrodynamical (MMH) simulations \cite{pen95,pen98}. We focus on
the angular power spectrum of the SZ effect and study its dependence
on cosmology. We compare our results to the Press-Schechter
predictions derived using the methods of KK99
. We study the redshift dependence of the SZ power
spectrum, and estimate the contribution of groups and filaments.  We
also study the effect of the finite resolution and finite box size of
the simulations. Results from projected maps of the SZ effect using
the same simulations are presented in Seljak et al.\cite{sel99}. 
We study the implications of our results for future and upcoming CMB
missions (see also Refs.\cite{agh97,ref99b}).

This paper is organized as follows. In \S\ref{sz}, we briefly describe
the SZ effect and derive expressions for the integrated comptonization
parameter and the SZ power spectrum.  In \S\ref{methods}, we describe our
different methods used to compute these quantities: hydrodynamical
simulations, the PS formalism, and a simple model with constant
bias. We present our results in \S\ref{results}, and discuss the
limitations imposed by the finite resolution and box size of the
simulations. Our conclusions are summarized in \S\ref{conclusions}.

\section{Sunyaev--Zel'dovich Effect}
\label{sz}

The SZ effect is produced from the inverse Compton scattering of CMB
photons \cite{sun72,sun80,rep95,bir99}. The resulting change in the
(thermodynamic) CMB temperature is
\begin{equation}
  \label{eq:dt_t}
  \frac{\Delta T}{T_{0}} = y j(x)
\end{equation}
where $T_{0}$ is the unperturbed CMB temperature, $y$ is the
comptonization parameter, and $j(x)$ is a spectral function defined in
terms of $x\equiv h \nu/k_B T_{0}$, $h$ is the Planck
constant and $k_B$ is the Boltzmann constant. In the nonrelativistic
regime, the spectral function is given by $j(x) = x (e^{x}+1)
(e^{x}-1)^{-1}-4$, which is negative (positive) for observation
frequencies $\nu$ below (above) $\nu_{0}\simeq 217$ GHz, for $T_{0}
\simeq 2.725$ K. In the Rayleigh-Jeans (RJ) limit ($x \ll 1$), $j(x)
\simeq -2$. The comptonization parameter is given by
\begin{equation}
  \label{eq:y_basic}
  y = \sigma_{T} \int dl~n_e\frac{k_B T_e}{m_{e}c^{2}}
    = \frac{\sigma_{T}}{m_{e}c^{2}} \int dl~p_{e}
\end{equation}
where $\sigma_{T}$ is the Thomson cross-section,
$n_e$, $T_e$ and $p_{e}$ are the
number density, temperature and thermal pressure of the electrons, 
respectively, and the integral is over the physical line-of-sight 
distance $dl$.

We consider a general FRW background cosmology with a scale
parameter defined as $a \equiv R/R_{0}$, where $R$ is the scale
radius at time $t$ and $R_{0}$ is its present value. The Friedmann
equation implies that $da = H_{0} \left( 1 - \Omega + \Omega_{m}
a^{-1} + \Omega_{\Lambda} a^{2} \right)^{1/2} dt$ where
$\Omega \equiv \Omega_{m} + \Omega_{\Lambda}$, $\Omega_{m}$, and
$\Omega_{\Lambda}$ are the present total, matter, and vacuum
density in units of the critical density $\rho_{c} \equiv 3
H_{0}^2/(8\pi G)$.  As usual, the Hubble constant today is
parametrized by $H_{0} \equiv 100~h$ km s$^{-1}$ Mpc$^{-1}$. It is
related to the present scale radius by $R_{0}=c/(\kappa H_{0})$,
where $\kappa^2 \equiv 1-\Omega$, 1, and $\Omega-1$ in a open,
flat, and closed cosmology, respectively. The comoving distance
$\chi$, the conformal time $\tau$, the light travel time $t$, and
the physical distance $l$ are then related by $dl = c dt = c a
d\tau = a d\chi$.

With these conventions, and assuming that the electrons and ions are
in thermal equilibrium, equation~(\ref{eq:y_basic}) becomes
\begin{equation}
  \label{eq:y_basic2}
  y = \sigma_{T}
      \int a d\chi \frac{\rho}{\mu_e m_p}
      \frac{k_B T}{m_e c^2}, 
\end{equation}
where $\rho$ is the gas mass density, $T$ 
is the gas temperature, and 
$\mu_{e}^{-1} \equiv n_e/(\rho/m_p)$
is the number of electrons per proton mass.
Equation~(\ref{eq:y_basic2}) can be written in
the convenient form
\begin{equation}
  \label{eq:y_dchi}
  y = y_{0}  \int d\chi T_{\rho} a^{-2},
\end{equation}
where 
$T_{\rho} \equiv \rho T/\overline{\rho}$ 
is the gas density-weighted temperature, and 
$\overline{\rho} = \rho_c \Omega_b a^{-3}$.
The overbar denotes a spatial average
and $\Omega_{b}$ is the present baryon density parameter.
The constant $y_0$ is given by 
\begin{eqnarray}
\label{eq:y0} 
  y_{0} 
  & \equiv & \frac{\sigma_{T} \rho_{c} \Omega_{b} k_B}
                  {\mu_{e} m_{p} m_{e} c^{2}} \nonumber \\ 
  & \simeq & 1.710\times 10^{-16} 
             \left( \frac{\Omega_bh^{2} }{0.05} \right) 
             \left( \frac{1.136}{\mu_{e}} \right) 
             \ \ {\rm K}^{-1}\ {\rm Mpc}^{-1},
\end{eqnarray}
where the central value for $\mu_{e}$ was chosen to correspond 
to a He fraction by mass of 0.24, and that for $\Omega_{b}$ to agree with 
Big Bang Nucleosynthesis constraints\cite{schramm98}.

The mean comptonization parameter $\overline{y}$ can be directly measured 
from the distortion of the CMB spectrum (see Ref. \onlinecite{ste97} 
for a review), and is given by
\begin{equation}
  \label{eq:y_bar}
  \overline{y} = y_{0}
  \int d\chi \overline{T}_{\rho} a^{-2}.
\end{equation}
It can thus be computed directly from the history of the volume-averaged
density-weighted temperature $\overline{T}_{\rho}$.  The gas in groups
and filaments is at a temperature of the order of $10^{7}\ {\rm K}$
(or $1\ {\rm keV}$), and thus induce a $y$-parameter of the order 
of $10^{-6}$ over a cosmological distance of $cH_{0}^{-1}\simeq
3000\ h^{-1}\ {\rm Mpc}$ (see Eq.~[\ref{eq:y0}]). 
This is one order of magnitude below the current
upper limit of $\overline{y} < 1.5 \times 10^{-5}$ (95\% CL) from the
COBE/FIRAS instrument \cite{fix96}.

The CMB temperature fluctuations produced by the SZ effect are
quantified by their spherical harmonics coefficients $a_{l m}$,
which are defined by $\Delta T ({\mathbf n}) = T_{0}^{-1}
\sum_{l m} a_{lm} Y_{l m}({\mathbf n})$. The angular
power spectrum of the SZ effect is then $C_{l} \equiv \langle
|a_{lm}|^{2} \rangle$, where the brackets denote an ensemble
average. Since most of the SZ fluctuations occur on small angular
scales, we can use the small angle approximation and consider the
Fourier coefficients $\widetilde{\Delta T}({\mathbf l}) = \int
d^{2}{\mathbf n} \Delta T({\mathbf n}) e^{i {\mathbf l} \cdot
{\mathbf n}}$. They are related to the power spectrum by $\langle
\widetilde{\Delta T}({\mathbf l}) \widetilde{\Delta
T}^{*}\!({\mathbf l}') \rangle \simeq T_{0}^{2} (2\pi)^{2}
\delta^{(2)}({\mathbf l}-{\mathbf l}') C_{l}$, where 
$\delta^{(2)}$
denotes the 2-dimensional Dirac-delta function. 
The SZ temperature variance is then
$\sigma_{\rm T}^{2} \equiv \left \langle \left( \Delta T/ T_{0}
\right)^{2} \right \rangle = \sum_{l} (2l+1) C_{l} / (4\pi)
\simeq \int \!dl \,l C_{l}/(2\pi)$. Since, as we will see,
$\overline{T}_{\rho} a^{-2}$ varies slowly in cosmic time scale and 
since the pressure fluctuations occur on scales much smaller than 
the horizon scale, we can apply Limber's equation in Fourier space 
(eg. Ref.~\onlinecite{kai98}) to equation~(\ref{eq:y_dchi}) and 
obtain,
\begin{equation}
\label{eq:cl} C_{l} \simeq j^2(x) y_{0}^{2} \int d\chi
\overline{T}_{\rho}^{2} P_{p}\!\left(\frac{l}{r},\chi
\right) a^{-4} r^{-2},
\end{equation}
where $r=R_{0}\sinh(\chi R_{0}^{-1})$, $\chi$, and $R_{0}\sin(\chi
R_{0}^{-1})$ are the comoving angular diameter distances in an open,
flat, and closed cosmology, respectively, and $P_{p}(k,\chi)$ is the
3-dimensional power spectrum of the pressure fluctuations, at a given
comoving distance $\chi$.  In general, we define the 3-dimensional
power spectrum $P_{q}(k)$ of a quantity $q$ by
\begin{equation}
\label{eq:p_q}
\langle \widetilde{\delta_{q}}({\mathbf k})
\widetilde{\delta_{q}}^{*}({\mathbf k'}) \rangle = (2\pi)^{3}
\delta^{(3)}({\mathbf k}-{\mathbf k'}) P_{q}(k),
\end{equation}
where $\widetilde{\delta_{q}}({\mathbf k}) = \int d^{3}x
\delta_{q}({\mathbf x}) e^{i {\mathbf k} \cdot {\mathbf x}} $, and
$\delta_{q} \equiv (q-\overline{q})/\overline{q}$. With these
conventions, the variance is $\sigma^{2}_{q} \equiv \langle
\delta_{q}^{2} \rangle = \int d^{3}k P_{q}(k)/(2\pi)^{3}$.  For a flat
universe, equation~(\ref{eq:cl}) agrees with the expression of Persi
et al. \cite{per95}. The SZ power spectrum can thus be readily
computed from the history of the mean density-weighted temperature
$\overline{T}_{\rho}(\chi)$ 
and of the pressure power spectrum $P_{p}(k,\chi)$.

\section{Methods}
\label{methods}

\subsection{Simulations}
\label{simulations}

We used the MMH code written by Pen\cite{pen95,pen98}, which was
developed by merging concepts from earlier hydrodynamic methods.
Grid-based algorithms feature low computational cost and high
resolution per grid element, but have difficulties providing the large
dynamic range in length scales necessary for cosmological
applications.  On the other hand, particle-based schemes, such as the
Smooth Particle Hydrodynamics (for a review see
Ref.\onlinecite{mon92}) fix their resolution in mass elements rather
than in space and are able to resolve dense regions.  However, due to
the development of shear and vorticity, the nearest neighbors of
particles change in time and must be determined dynamically at each
time step at a large computational cost.

To resolve these problems, several approaches have recently been
suggested \cite{sha96,gne95,bry94}. The MMH code combines the
advantages of both the particle and grid-based approaches by deforming
a grid mesh along potential flow lines.  It provides a twenty fold
increase in resolution over previous Cartesian grid Eulerian schemes,
while maintaining regular grid conditions everywhere \cite{pen98}.
The grid is structured in a way that allows the use of high resolution
shock capturing TVD schemes (see for example Ref. \onlinecite{kan94}
and references therein.) at a low computational cost per grid cell.
The code that optimized for parallel processing, which is
straightforward due to the regular mesh structure. The moving mesh
provides linear compression factors of about 10, which correspond
to compression factors of about $10^{3}$ in density. Note that this
code does not include the effects of cooling and feedback of the gas.

We ran three simulation with $128^{3}$ curvilinear cells, corresponding
to $\sigma_{8}$-normalized SCDM, $\Lambda$CDM, and OCDM models. The
simulation parameters are listed in table~\ref{tab:simulations}.
Note that in all cases, the shape parameter for the linear power
spectrum was set to $\Gamma=\Omega_{m} h$ \cite{pea96,sug95}.
The simulation output was saved at $z=0,0.5,1,2,4,8$ and $16$,
and was used to compute 3-dimensional statistics.

To test the resolution of the simulation, we compared the power
spectrum of the dark matter density fluctuations $P_{\rho DM}(k)$
(defined in Eq.~[\ref{eq:p_q}] with $q \equiv \rho_{DM}$) from the simulations
to that from the Peacock \& Dodds \cite{pea96} fitting formula. The
results for the $\Lambda$CDM are shown on figure~\ref{fig:pk_rhodm},
and are similar for the other three models. The simulation power
spectrum agrees well with the fitting formula for $0.2 \lesssim k
\lesssim 2\ h$ Mpc$^{-1}$ at all redshifts. For $k \lesssim 0.2$ and $k
\gtrsim 2\ h$ Mpc$^{-1}$, the simulations are limited by the finite
size of the box and the finite resolution, respectively. We will use
these limits below, to study the effect of these limitations on the SZ
power spectrum.
\label{resolution}

\subsection{Press--Schechter Formalism}
\label{ps}

It is useful to compare the simulation results with analytic
calculations based on the Press--Schechter (PS)
formalism\cite{pre74}. We compute the angular power spectrum and the
mean Comptonization parameter, using the methods of KK99
and Barbosa et al.\cite{bar96}, respectively.
For definitiveness, we adopt the spherical isothermal $\beta$ model
with the gaussian-like filter for the gas density distribution in a cluster,
\begin{equation}
  \label{eq:betamodel}
   \rho_{\rm gas}(r) 
 = \rho_{\rm gas0}
   \left[
         1 + \left(\frac{r}{r_c}\right)^2
   \right]^{-3\beta/2}
   e^{-r^2/\xi R^2},
\end{equation}
where $R$ and $r_c$ are the virial radius and the core radius of a
cluster, respectively, and a fudge factor $\xi=4/\pi$ is taken to
properly normalize the gas mass enclosed in a cluster\cite{kom99}.  We
employed a self-similar model for the cluster
evolution\cite{Kaiser86}.  Note that other evolution models yield
spectra that differ only at small angular scales
($l>2000$)\cite{kom99}.
The gas mass fraction of objects is taken to be the cosmological
mean, i.e., $\Omega_b/\Omega_m$.

The volume-averaged density-weighted temperature is given by
\begin{equation}
  \label{eq:PSTrho}
   \overline{T}_\rho(z)
   = \frac1{\overline{\rho}_0}
     \int_{M_{\rm min}}^{M_{\rm max}} dM M\frac{dn(M,z)}{dM} 
                                         T(M,z),
\end{equation}
where $\overline{\rho}_0=2.775\ \Omega_{m}\ h^2\ M_\odot\ {\rm Mpc^{-3}}$ is
the present mean mass density of the universe, $dn/dM$ is the PS mass
function which gives the comoving number density of collapsed objects
of mass $M$ at $z$. $T$ is computed by the virial temperature 
given by
\begin{eqnarray}
   \label{eq:Tvir}
   k_B T(M,z) 
      & = & 5.2\ \beta^{-1}\left(\frac{\Delta_c(z)}{18\pi^2}\right)^{1/3}
	    \left(\frac{M}{10^{15}\ h^{-1}\ M_\odot}\right)^{2/3} 
            \nonumber \\
	& & \times (1+z)\ \Omega_m^{1/3}\ {\rm keV},
\end{eqnarray}
where $\Delta_c(z)$ is the mean mass density of a collapsed object at
$z$ in units of $\overline{\rho}_0\Omega_m(1+z)^3$\cite{LC93,NS97}. 
While Barbosa et al.\cite{bar96} used $\beta\simeq 5/6$, 
we adopt $\beta=2/3$ according to KK99.  

The limits $M_{\rm min}$ and $M_{\rm max}$ should be taken to fit the
resolved mass range in the simulation. The mass enclosed in the
spherical top-hat filter with comoving wavenumber $k$ is
\begin{eqnarray}
   \label{eq:mass}
   M & = & \frac{4\pi}3\overline{\rho}_0\left(\frac{\pi}k\right)^3 \nonumber \\
     & = & 3.6\times 10^{13}\left(\frac{k}{1\ h\ {\rm Mpc^{-1}}}\right)^{-3}
           \Omega_{m} \ h^{-1}\ M_\odot.	
\end{eqnarray}
Since the $k$-range of confidence in the simulation is 
approximately $0.2 \lesssim k
\lesssim 2\ h\ {\rm Mpc^{-1}}$ (see \S\ref{simulations}), equation
(\ref{eq:mass}) gives $M_{\rm min} \simeq 4.5\times 10^{12}\ \Omega_m\
h^{-1}\ M_\odot$ and $M_{\rm max} \simeq 4.5\times 10^{15}\ \Omega_m\
h^{-1}\ M_\odot$.  This mass range is used for calculating the angular
power spectrum, the mean Comptonization parameter, and the
density-weighted temperature. A more detailed inspection of
Figure~\ref{fig:pk_rhodm} reveals that the resolution of the
simulations depends on redshift, and involves a power law cutoff in
$k$ rather than a sharp cutoff. This must be kept in mind when
comparing the two methods (see \S\ref{limitations}).

\subsection{Constant Bias Model}
\label{bias_model}
It is also useful to consider a simple model with constant bias.  The
bias $b_{p}$ of the pressure with respect to the DM density can be
defined as
\begin{equation}
\label{eq:bp}
b_{p}^{2}(k,z) \equiv \frac{P_{p}(k,z)}{P_{\rho_{DM}}(k,z)},
\end{equation}
and generally depends both on wave number $k$ and redshift $z$.  In
this simple model, we assume that $b_{p}$ is independent of both $k$
and $z$, and replace the pressure power spectrum $P_{p}(k,z)$ in
Equation~(\ref{eq:cl}), by $b_{p} P_{\rho DM}(k,z)$, where $P_{\rho DM}$
is evaluated using the Peacock \& Dodds fitting formula \cite{pea96}.
This has the advantage of allowing us to extend the contribution
to the SZ power spectrum to arbitrary ranges of $k$. This will be
used in \S\ref{limitations} to test the effect of finite resolution
and finite box size on the SZ power spectrum.

\section{Results}
\label{results}

\subsection{Projected Maps}

\begin{figure}
\vspace*{1.2cm}
\centerline{\epsfig{file=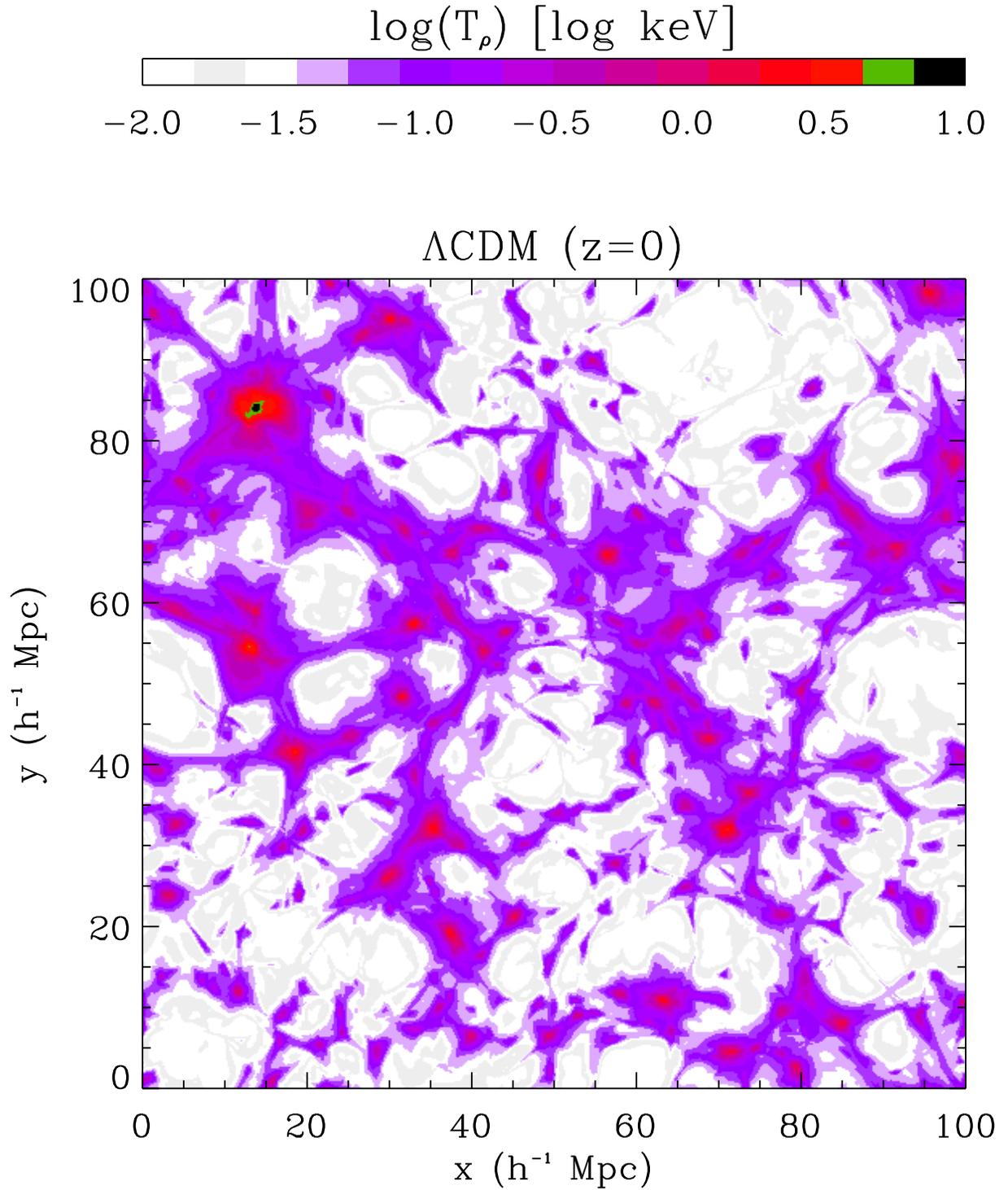}}
\caption{Density-weighted temperature for the
$\Lambda$CDM simulation at $z=0$. The temperature map
was derived by projecting through one face of the
box.}
\label{fig:trho}
\end{figure}

\begin{figure}
\vspace*{1.2cm}
\centerline{\epsfig{file=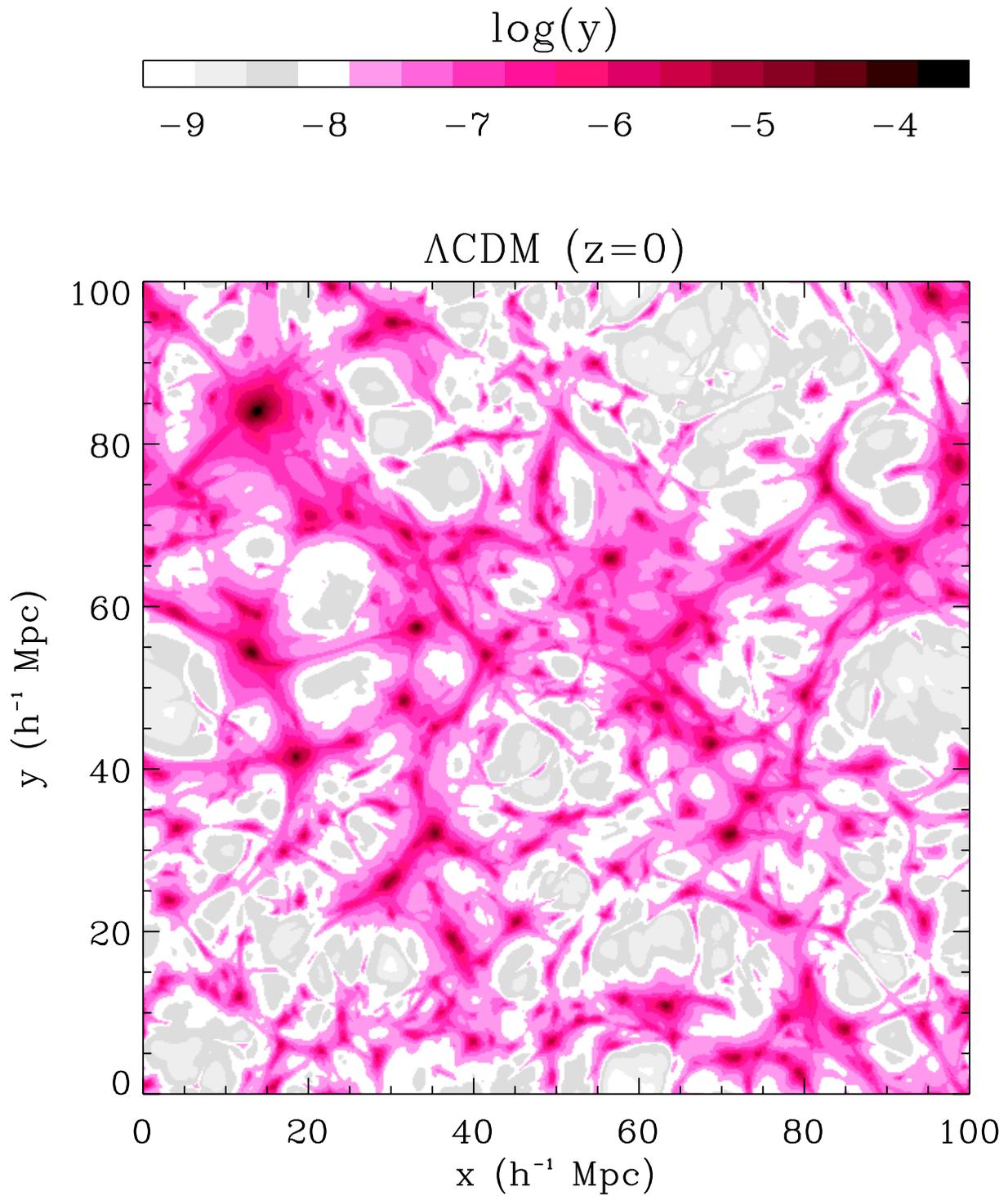}}
\caption{Comptonization-parameter map for the $\Lambda$CDM simulation
at $z=0$.}
\label{fig:y}
\end{figure}

Figure~\ref{fig:trho} shows a map of the density-weighted temperature
for the $\Lambda$CDM model projected through one box at
$z=0$. Clusters of galaxies are clearly apparent as regions with $k_B T
\gtrsim 3$ keV. The gas in filaments and groups can be seen to stretch
between clusters and has temperatures in the range $0.1\lesssim k_B T
\lesssim 3$ keV. While these regions have smaller temperatures, they
have a relatively large covering factor and can thus contribute
considerably to the $y$-parameter and to the SZ fluctuations.  This
can be seen more clearly in Figure~\ref{fig:y}, which shows the
corresponding map of the comptonization parameter. Clusters produce
$y$-parameters greater than $10^{-5}$, while groups and filaments
produce $y$-parameters in the range $10^{-7} - 10^{-5}$. Note that of
the total SZ effect on the sky would include contributions for a
number of simulation boxes along the line-of-sight. In such a map, the
filamentary structure is less apparent as filaments are averaged out
by projection \cite{das99,sel99}. A quantitative analysis of the
contribution of groups and filaments to the SZ effect is presented in
the following sections.

\subsection{Mean Comptonization Parameter}
\label{y_bar}

\begin{figure}
\vspace*{3cm}
\centerline{\epsfig{file=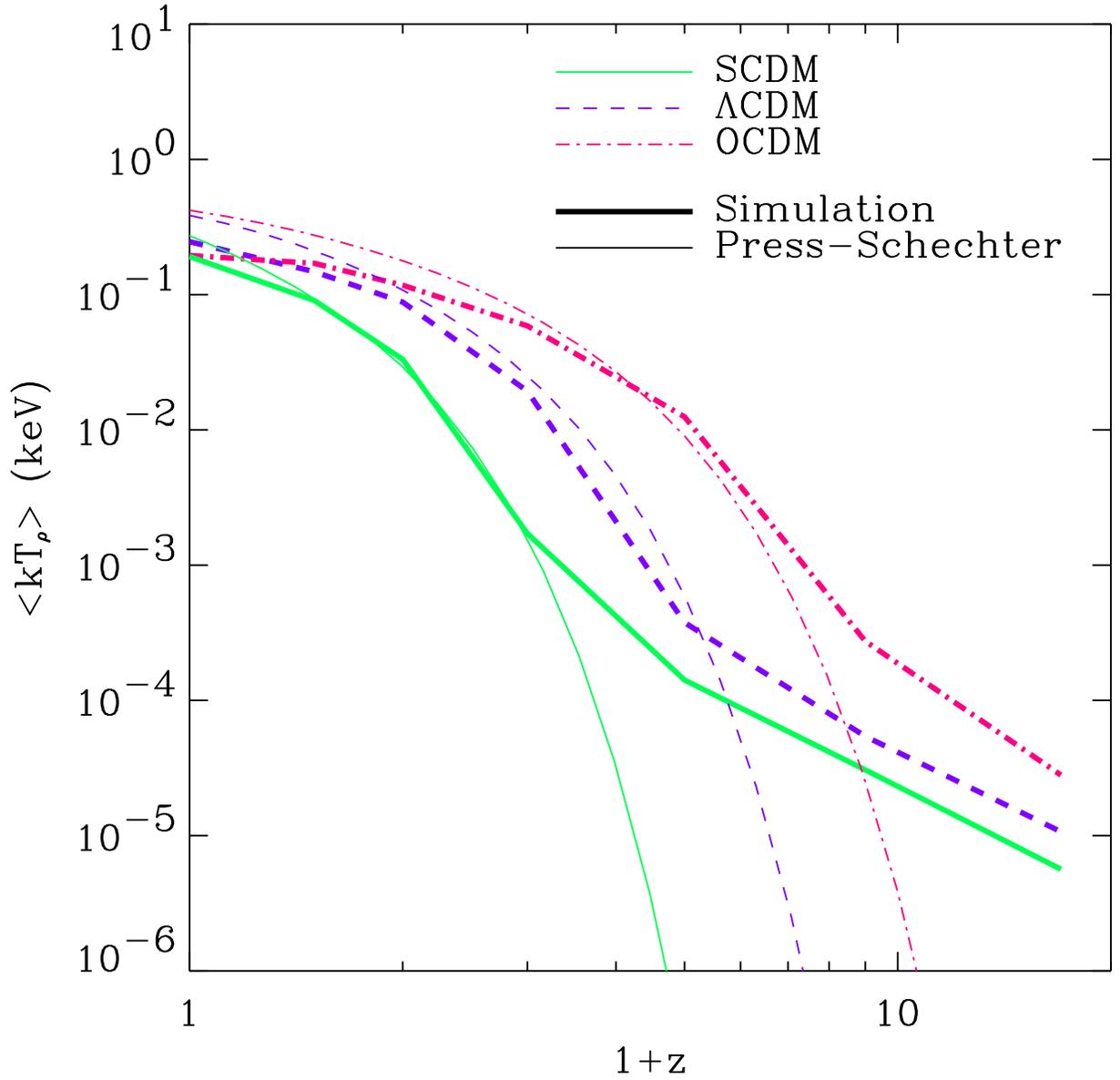}}
\caption{Temperature history of the gas. For each model, the
density weighted temperature $T_{\rho}$ is shown for the simulations
and for the Press-Schechter prediction.}
\label{fig:t_z}
\end{figure}

The evolution of the density-weighted temperature for each of the
simulations is shown on figure~\ref{fig:t_z}. The temperatures at
present are listed in table~\ref{tab:results} and are quite similar. This
is expected since all models were chosen to have similar $\sigma_{8}$
normalizations. The evolution is steeper for the SCDM, flatter for the
OCDM model, and intermediate for the $\Lambda$CDM model. This is
consistent with the different rate of growth of structure for each
model.

Also plotted on this figure is the density-weighted temperature
derived from the PS formalism (Eq.~[\ref{eq:PSTrho}]). The agreement
for $z\lesssim 4$ is good, both for the relative amplitudes and for
the shapes of the temperature evolution. At $z\gtrsim 4$ the
non-linear mass scale is not sufficiently large compared to the mass
resolution of the simulation, so the temperatures are not meaningful
in that regime.  This is however not a serious limitation, as these
redshifts do not contribute significantly to either the mean
comptonization or the SZ fluctuations. The PS temperatures exceed the
simulations at low redshift for all cosmological models, since massive
(high temperature) clusters, which may be missed in the simulations
due to the effect of finite box size, dominate there. The PS
temperatures at $z=0$ are listed in Table~\ref{tab:results}.

The parameters of our $\Lambda$CDM model were chosen to coincide with
that for the simulation of Cen \& Ostriker\cite{cen99}. While the
slope of our density-weighted temperature agrees approximately with
theirs for $z\lesssim 3$, the amplitude is significantly
different. They find a final temperature of about 0.9 keV, which is a
factor of about 5 larger than ours. This discrepancy could be due to
the fact that their simulation include feedback from star formation,
while ours only comprise gravitational forces. It is however
surprising that standard feedback could produce such a large
difference.  One can estimate the gravitational binding energy of
virialized matter from the cosmic energy equation
\cite{pen99,davis97} and finds a thermal component of fluids to be
of order 1/4 keV, consistent with our simulations.  We should note,
however, that the high thermal temperatures from feedback may be
required for consistency with the X-ray background constraints\cite{pen99}.
The reason for this discrepancy is still unknown at
present, but should be kept in mind for the interpretation of our
results.

\begin{figure}
\vspace*{1.2cm}
\centerline{\epsfig{file=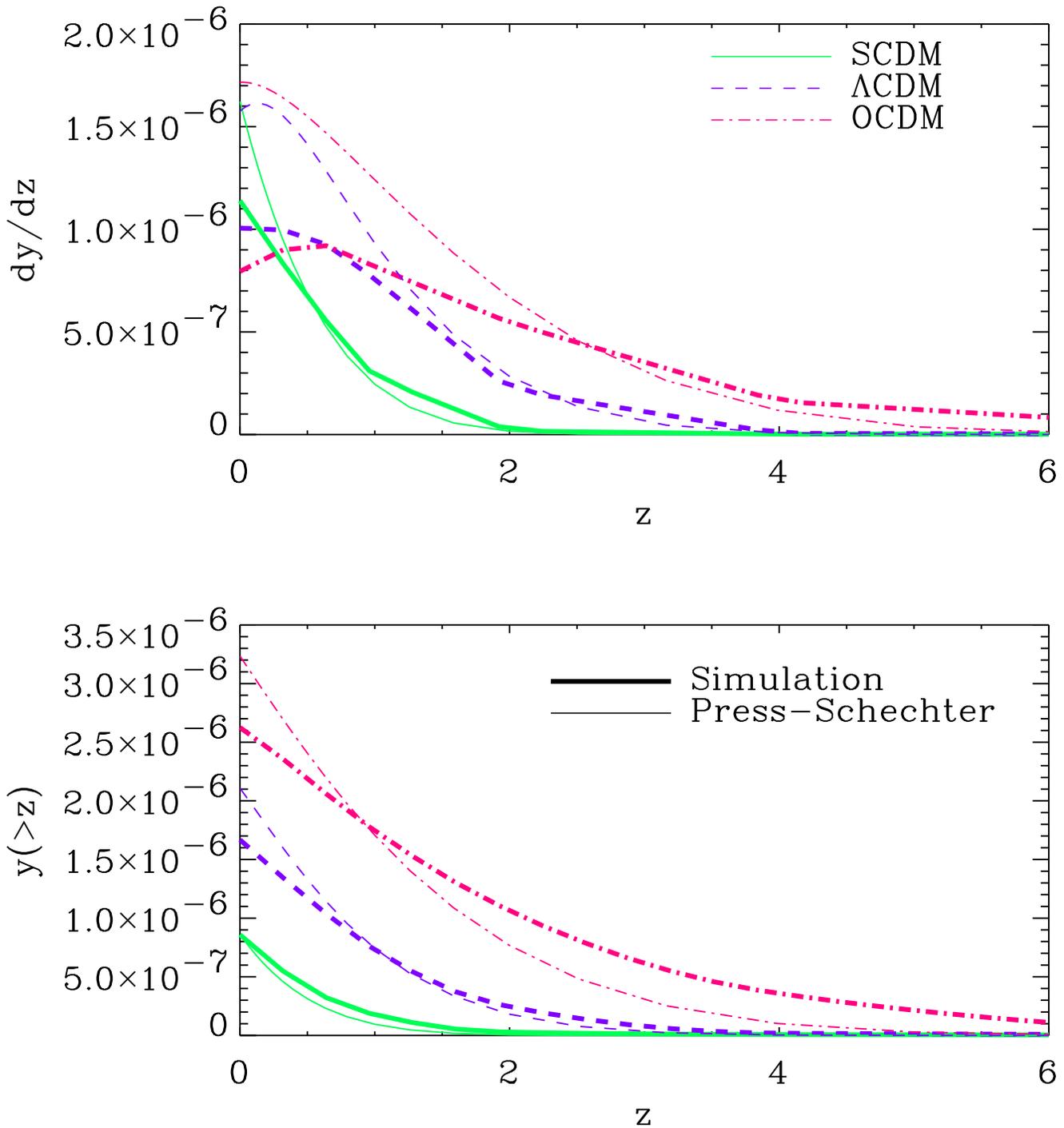}}
\caption{Differential and cumulative $y$-parameter versus redshift,
for each simulation. The cummulative $y$-parameter predicted from the
Press-Schechter formalism is shown as the thin lines.}
\label{fig:y_z}
\end{figure}

The mean comptonization parameter for each simulation was derived
using equation~(\ref{eq:y_bar}) and is listed in
table~\ref{tab:results}. In all cases, $\overline{y}$ is well below the
upper limit $\overline{y} < 1.5 \times 10^{-5}$ (95\% CL) set by the
COBE/FIRAS instrument \cite{fix96}. The differential and cumulative
redshift dependence of $\overline{y}$ are shown on
figure~\ref{fig:y_z}. For the three models, most of the mean SZ effect
is produced at $z\lesssim2$. The contribution from high redshift is
largest for the OCDM and smallest for the SCDM model, again in
agreement with the relative growth of fluctuations in each model. 

The differential and cumulative redshift dependence of $\overline{y}$
derived from the PS formalism are shown on this figure as the thin
lines. The values of $\overline{y}$ from PS are also listed in
Table~\ref{tab:results}. They are higher than that for the simulations
by about 25\% for the $\Lambda$CDM and OCDM models, are in close
agreement for the SCDM model.  The shapes of the differential curves
approximately agree, although the PS formalism predicts more
contributions from lower redshifts. This is can be traced
to the slightly steeper evolution of the PS temperatures in
figure~\ref{fig:t_z}, and is due to massive nearby clusters.

\subsection{Power Spectrum}
\label{cl}

\begin{figure}
\vspace*{3cm}
\centerline{\epsfig{file=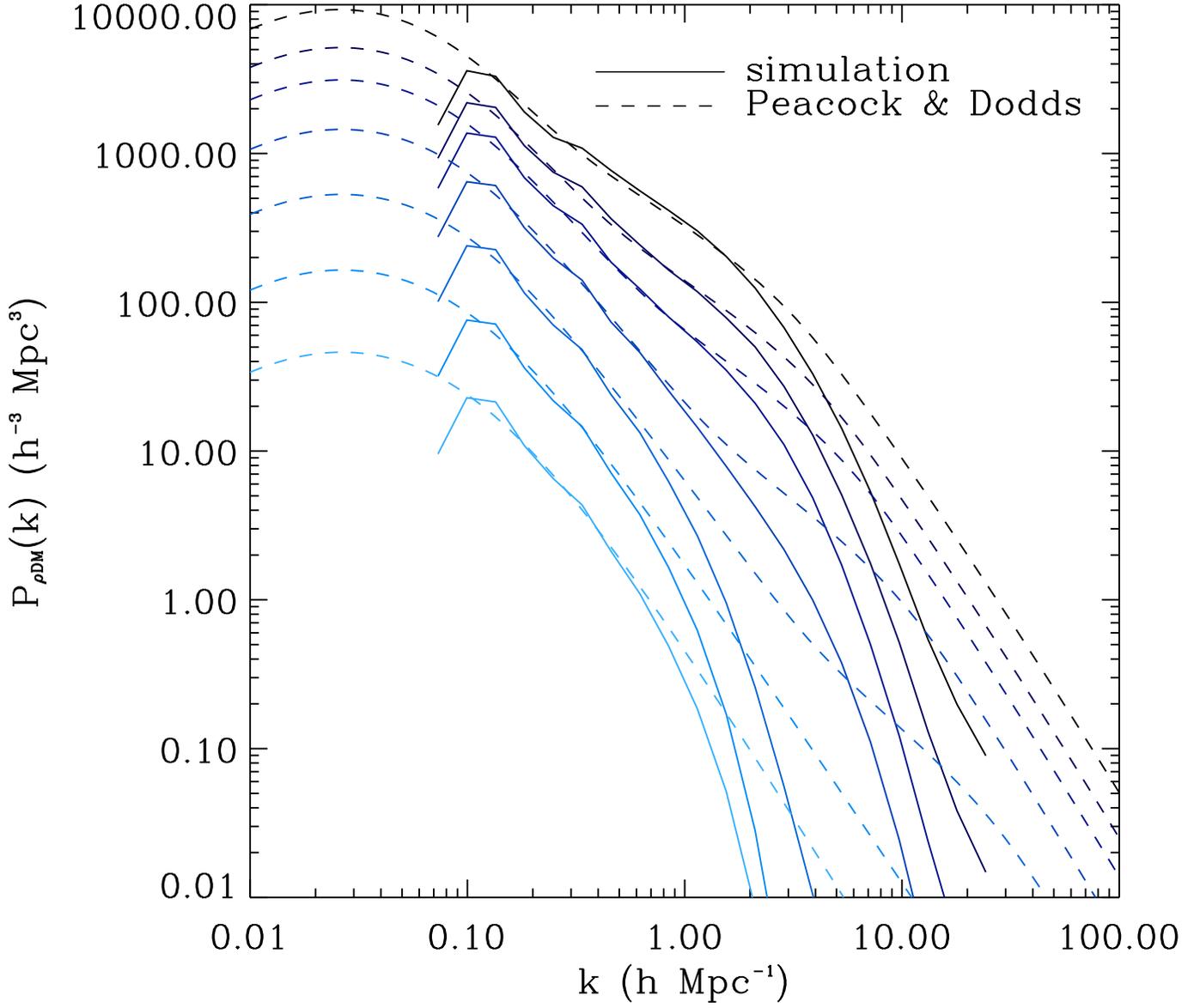}}
\caption{Power Spectrum of the DM density for the $\Lambda$CDM
simulation. The power spectrum from the simulation (solid
lines) is compared to that from the Peacock \& Dodds (1996)
fitting formula (dashed lines). The spectra, from top to bottom,
correspond to $z$=0, 0.5, 1, 2, 4, 8, and 16.}
\label{fig:pk_rhodm}
\end{figure}

\begin{figure}
\vspace*{3cm}
\centerline{\epsfig{file=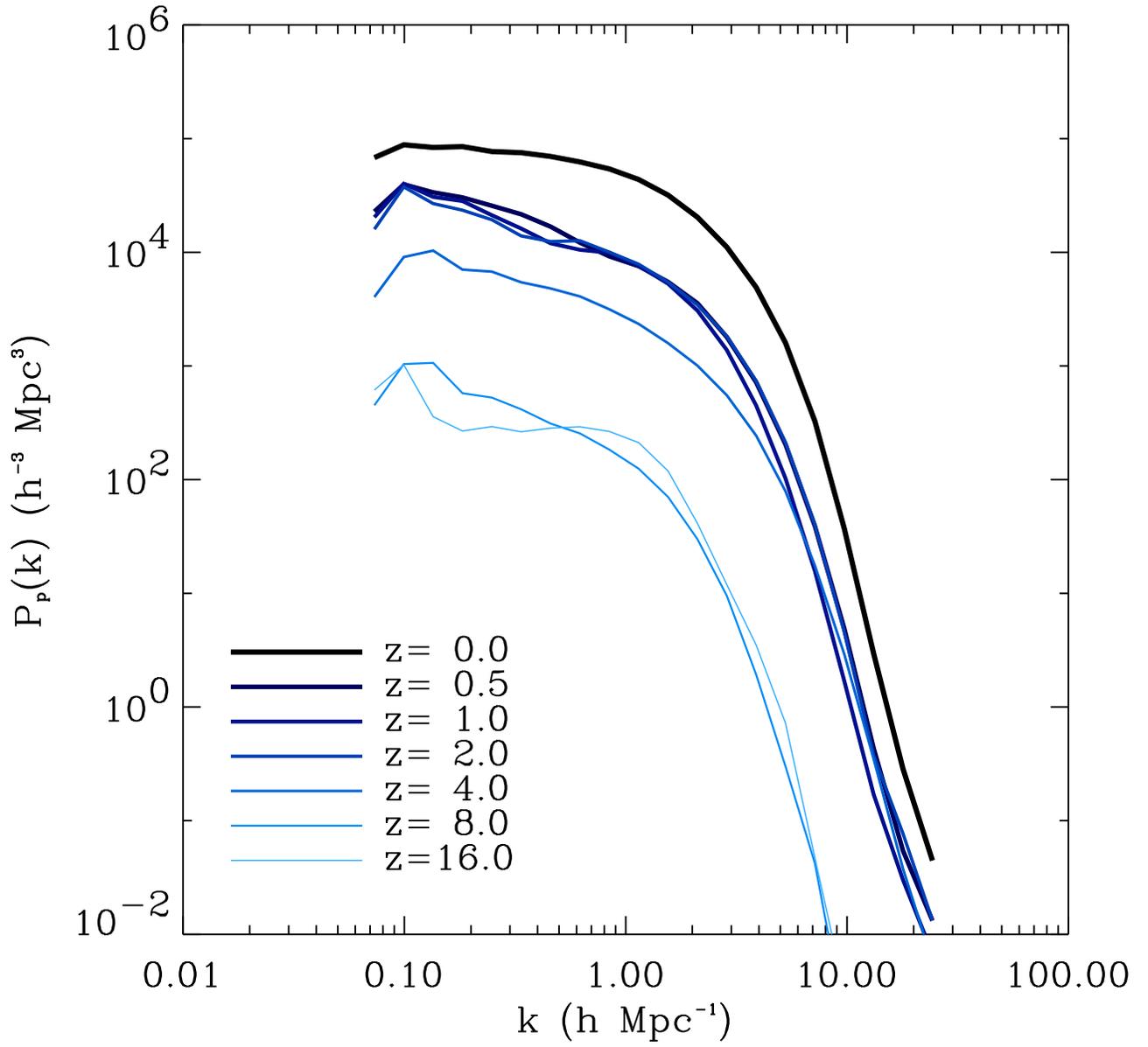}}
\caption{Power spectrum of the pressure fluctuations for the
$\Lambda$CDM model.}
\label{fig:pk_p}
\end{figure}

\begin{figure}
\vspace*{3cm}
\centerline{\epsfig{file=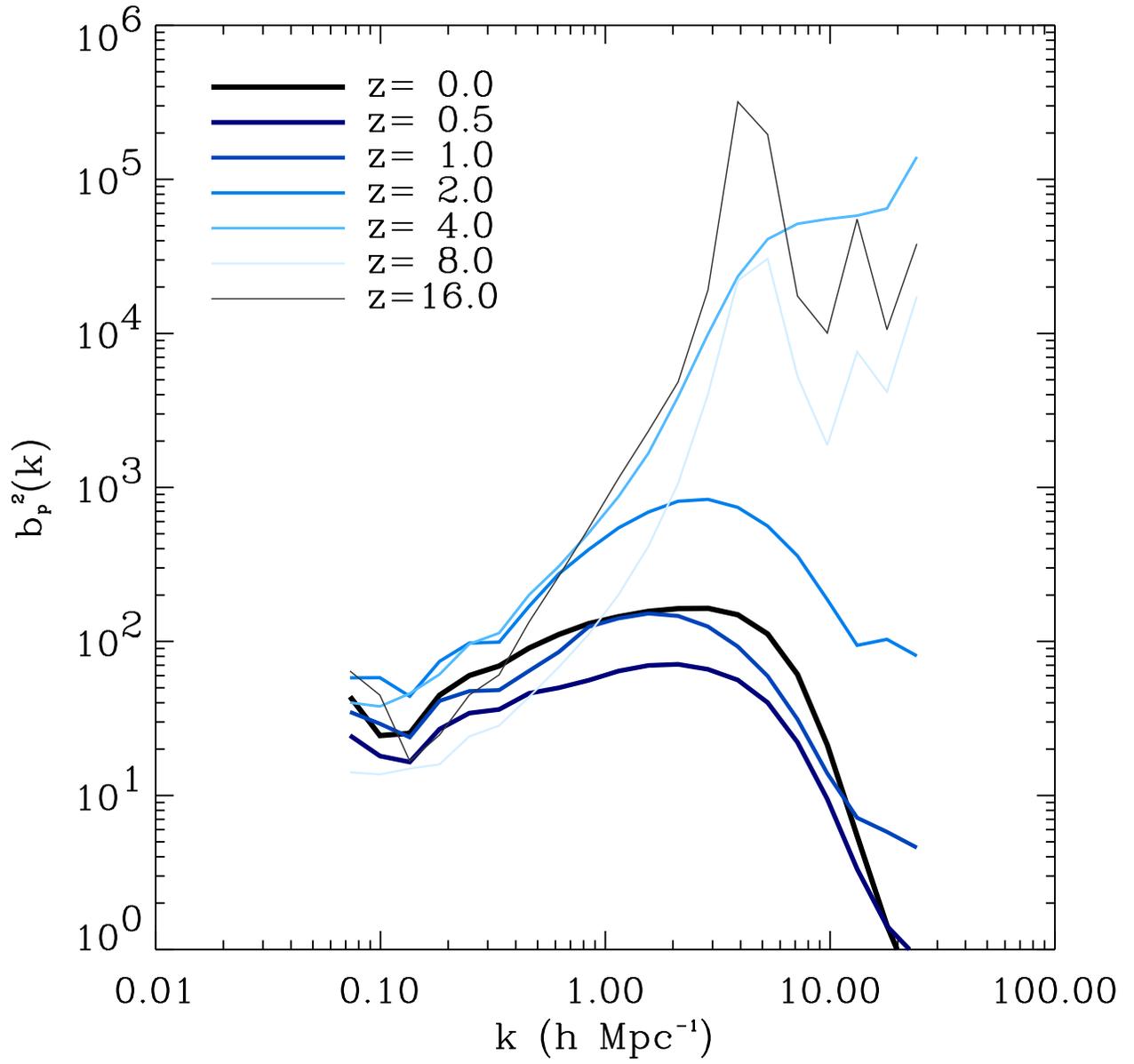}}
\caption{Bias of the pressure $b_{p}(k,z)$ for the $\Lambda$CDM
model.}
\label{fig:bp}
\end{figure}

As noted in \S\ref{sz}, the SZ power spectrum can be derived from the
history of the temperature $\overline{T}_{\rho}$ and of the pressure power
spectrum $P_{p}(k)$ (Eq.~[\ref{eq:cl}]). The evolution of the pressure
power spectrum is shown on figure~\ref{fig:pk_p}, for the $\Lambda$CDM
simulation. The amplitude of $P_{p}(k)$ increases with redshift, while
keeping an approximately similar shape. Perhaps more instructive is
the evolution of the pressure bias $b_{p}(k,z)$ (Eq~[\ref{eq:bp}]),
which is shown on figure~\ref{fig:bp}. For $z\lesssim 1$, $b_{p}$ is
approximately independent of scale, in the $k$-range of confidence
($0.2\lesssim k\lesssim2\ h$ Mpc$^{-1}$; see figure~\ref{fig:pk_rhodm}).
The value of $b_{p}$ at $z=0$ and $k=0.5\ h$ Mpc$^{-1}$ is listed in
table~\ref{tab:results}. For $z \gtrsim 1$, $b_{p}$ remains approximately
constant on large scales, but is larger on small scales. Indeed, at
early times, only a small number of small regions have collapsed and
are thus sufficiently hot to contribute to the pressure. As a result,
the pressure at high-$z$ is more strongly biased on small scales. 

The SZ angular power spectrum derived from integrating the pressure
power spectrum along the line of sight (Eq.~[\ref{eq:cl}]) is shown in
Figure~\ref{fig:cl} for each simulation. For comparison, the spectrum
of primary CMB anisotropies was computed using CMBFAST\cite{zal99},
and was also plotted on this figure as the solid line.  The SZ power
spectrum can be seen to be two order of magnitude below the primordial
power spectrum below $l \lesssim 2000$, but comparable to it beyond
that. Because of finite resolution and box size, the SZ power spectra
should be interpreted as lower limits outside of the $l$-range of
confidence highlighted by thicker lines (see \S\ref{limitations}).
The SCDM spectrum is lower than that for the $\Lambda$CDM and
OCDM. This is a consequence of the lower value of $\sigma_{8}$ for
this model. Indeed, KK99 have shown that the SZ power spectrum scales
as $C_l \propto \Omega_b^2 \sigma_8^6 h$, and is thus very sensitive
on this normalization. This scaling relation also allows us to compare
our results to that of the SCDM calculation of Persi et
al.\cite{per95}. The amplitude of their power spectrum, rescaled to the
same value of $\sigma_{8}$, is within 20\% of ours at $l=1000$, while
its shape is similar to ours.

\begin{figure}
\vspace*{1.2cm}
\centerline{\epsfig{file=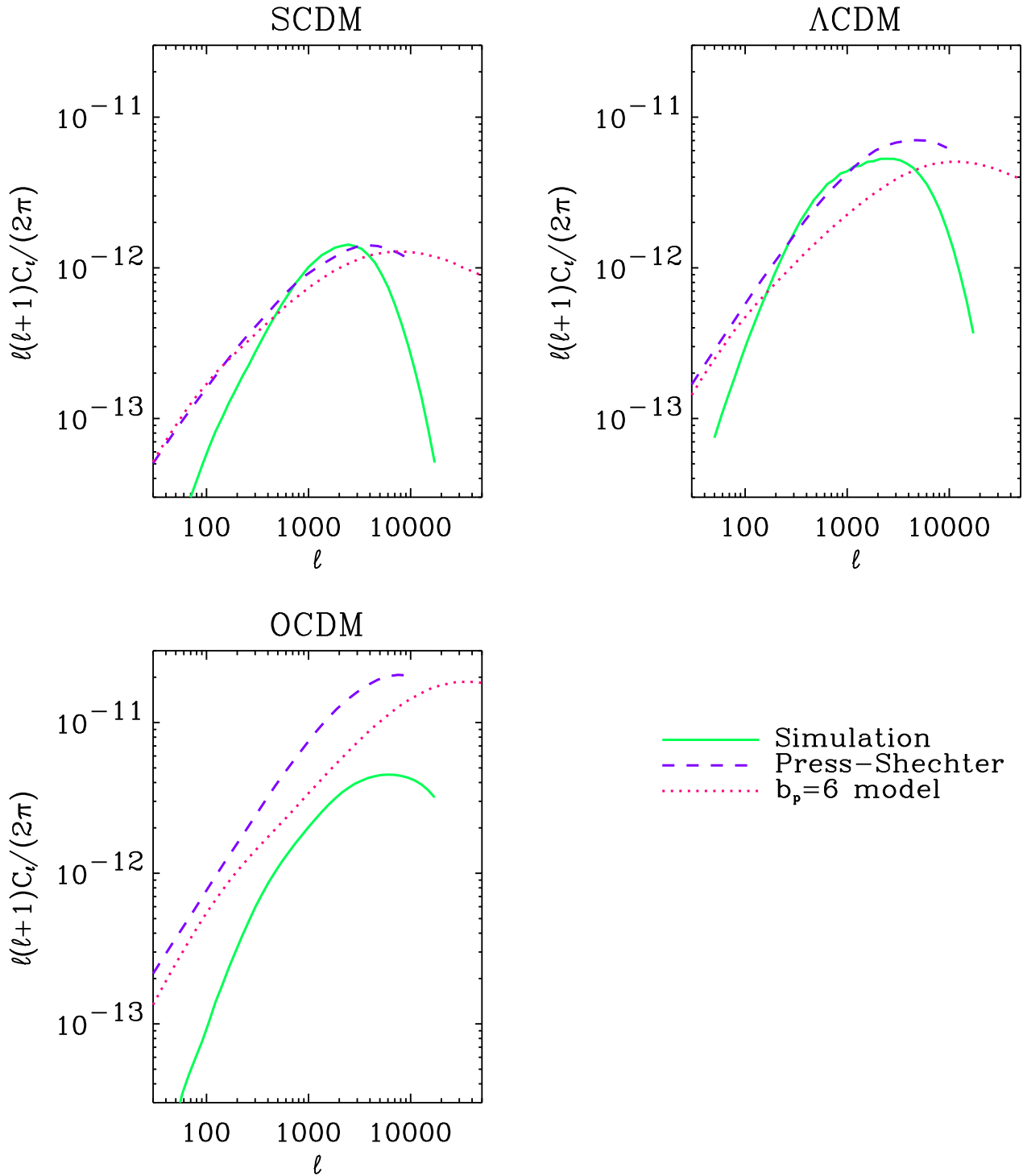}}
\caption{Angular power spectrum of the SZ effect from each method:
simulations, the PS formalism, and the constant bias model (with
$b_{p}=6$). These results are shown for the RJ regime, for
each cosmological models.}
\label{fig:cl_methods}
\end{figure}

Figure~\ref{fig:cl_methods} presents a comparison of the SZ power
spectra derived from each of the three methods described in
\S\ref{methods}.  For both the simulations and the PS formalism, the
SZ power spectra peak around $l\simeq 2000$ for the SCDM and
$\Lambda$CDM models, and around $l \simeq 5000$ for the OCDM model. On
the other hand, the constant bias models, which do not have a mass or
$k$ cutoff, peak at $l \simeq 10000-30000$. This is explained by the
fact that this model does not have a mass or $k$ cutoff and has
therefore more power on small scales. In \S\ref{limitations}, we will
use this comparison to study the effect of finite resolution and box
size of the simulations.

For $200 \lesssim l \lesssim 2000$, the simulation and PS predictions
approximately agree for the SCDM and $\Lambda$CDM models.  On the
other hand, for the OCDM model, the PS prediction is a factor of 3
higher than that from the simulations in this range.  This can be
traced to the fact that the $\Lambda$CDM simulation yields a larger
pressure bias $b_{p}$ (Eq.~\ref{eq:bp}) at low redshifts than the OCDM
simulation. By inspecting the figure corresponding to
Figure~\ref{fig:pk_rhodm} for the OCDM model, we indeed noticed that
more power was missing on small scales in this simulation. This is
probably due to the fact that the OCDM simulation was started at a
higher redshift ($z=100$) than the other two simulations ($z=30$).
Due to truncation errors in the Laplacian and gradient calculations,
modes with frequencies close to the Nyquist frequency are known to
grow much more slowly even in the linear regime. This effect is
reduced if the simulation is started later.

\begin{figure}
\vspace*{3cm}
\centerline{\epsfig{file=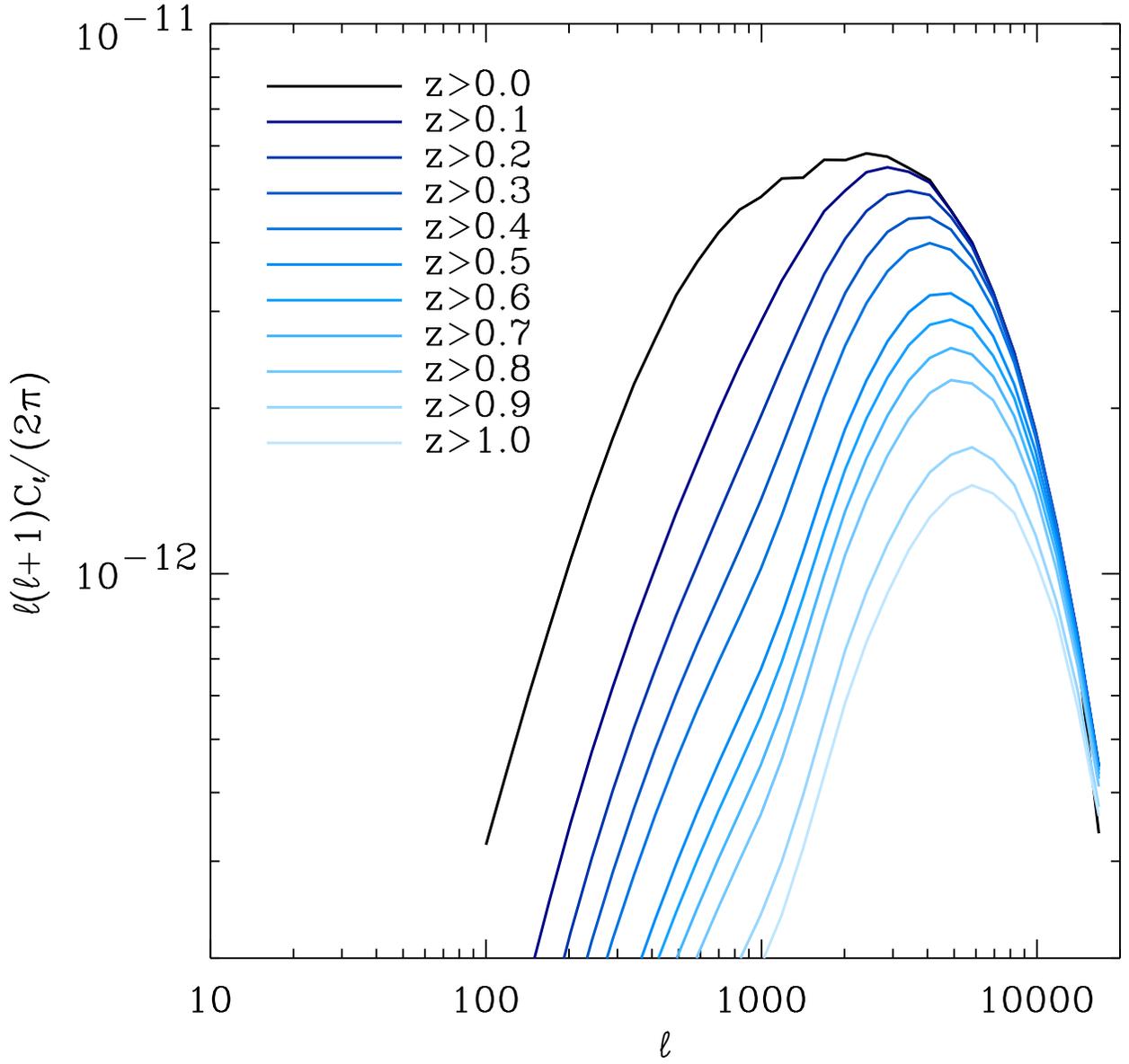}}
\caption{Dependence of SZ power spectrum on redshift. For the
$\Lambda$CDM model, the contribution to the SZ power spectrum up to
a given redshift is shown.}
\label{fig:cl_z}
\end{figure}

\begin{figure}
\vspace*{3cm}
\centerline{\epsfig{file=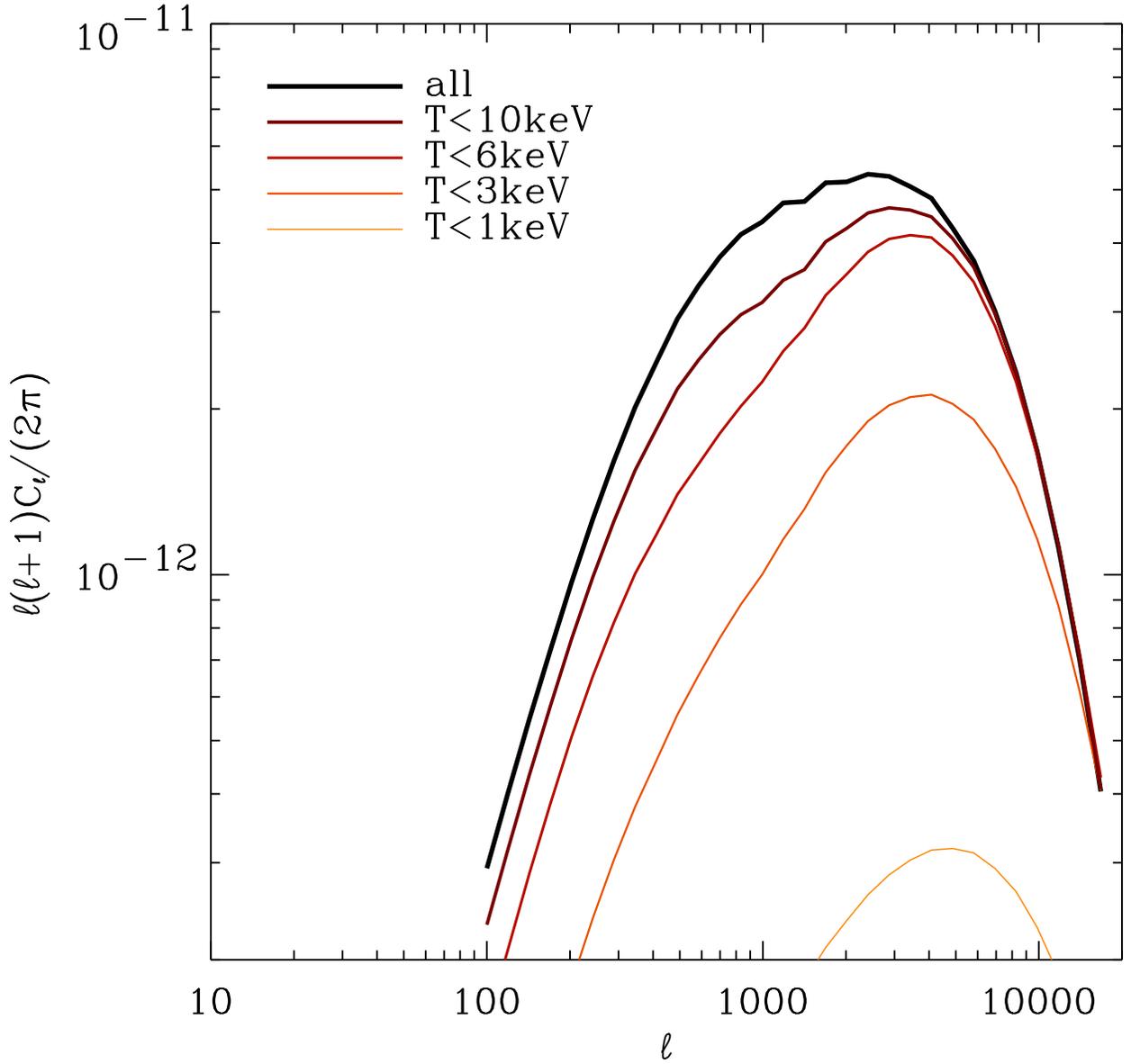}}
\caption{SZ power spectrum as a function of temperature for
the $\Lambda$CDM model. The SZ power spectrum was calculated after
removing regions with temperatures above the specified cutoff.}
\label{fig:cl_t}
\end{figure}

The redshift dependence of the SZ power spectrum is shown in
figure~\ref{fig:cl_z}, for the $\Lambda$CDM model. Most of the SZ
fluctuations are produced at low redshifts: at $l=500$, about 50\% of
the power spectrum is produced at $z\lesssim 0.1$, and about 90\% at
$z\lesssim 0.5$. The contribution of the warm gas in groups and
filaments can be studied by examining figure~\ref{fig:cl_t}. This
figure shows the $\Lambda$CDM power spectrum measured after removing
hot regions from the simulation volume, for several cutoff
temperatures. Approximately 50\% of the SZ power spectrum at $l=500$
is produced by gas with $k_B T \lesssim 5\ {\rm keV}$. In
\S\ref{prospects}, we show that these combined facts give good
prospects for the removal (and the detection) of SZ fluctuations from
CMB maps.

The behavior of the power spectrum for $l\lesssim 1000$, in
figures~\ref{fig:cl_t} and \ref{fig:cl_z}, agrees with the results of
KK99 who studied the Poisson and clustering contributions separately.
At low $l$'s , the SZ power spectrum is produced primarily by bright
(low-redshift or high-temperature) objects, i.e., by massive clusters,
and is thus dominated by the Poisson term.  However, after subtracting
bright clusters from the SZ map, the correlation term dominates the
Poisson term at high redshift.  Therefore, the SZ spectrum on large
angular, measured after subtracting bright spots, should trace
clustering at high redshift. This interesting effect will be discussed
in details elsewhere.

\subsection{Limitations of the Simulations}
\label{limitations}

It is important to assess the effect of the limitations of the
simulations on these results. First, the finite resolution may lower
the temperature $T_{\rho}$, since it prevents small scale structures
from collapsing. As we saw in \S\ref{ps}, the resolution limits of the
simulations correspond to halo masses of about $4.5 \times 10^{12}\
\Omega_m\ h^{-1}\ M_{\odot}$.  According to the PS formalism, the
contribution to $T_{\rho}$ from halos with masses smaller than this
limit is about $0.01-0.02$ keV, for $z\lesssim 4$ in the $\Lambda$CDM
model.  The SZ power spectrum at $l \lesssim 2000$ is produced
mainly at low redshifts, and is therefore little affected by this
effect. Note however, that $\overline{y}$, which is sensitive to small
halos at high redshifts, is more affected. Indeed, the
contribution to $\overline{y}$ by these halos is about $0.70\times
10^{-6}$, assuming a gas mass fraction of $\Omega_b/\Omega_m$.

\begin{figure}
\vspace*{3cm}
\centerline{\epsfig{file=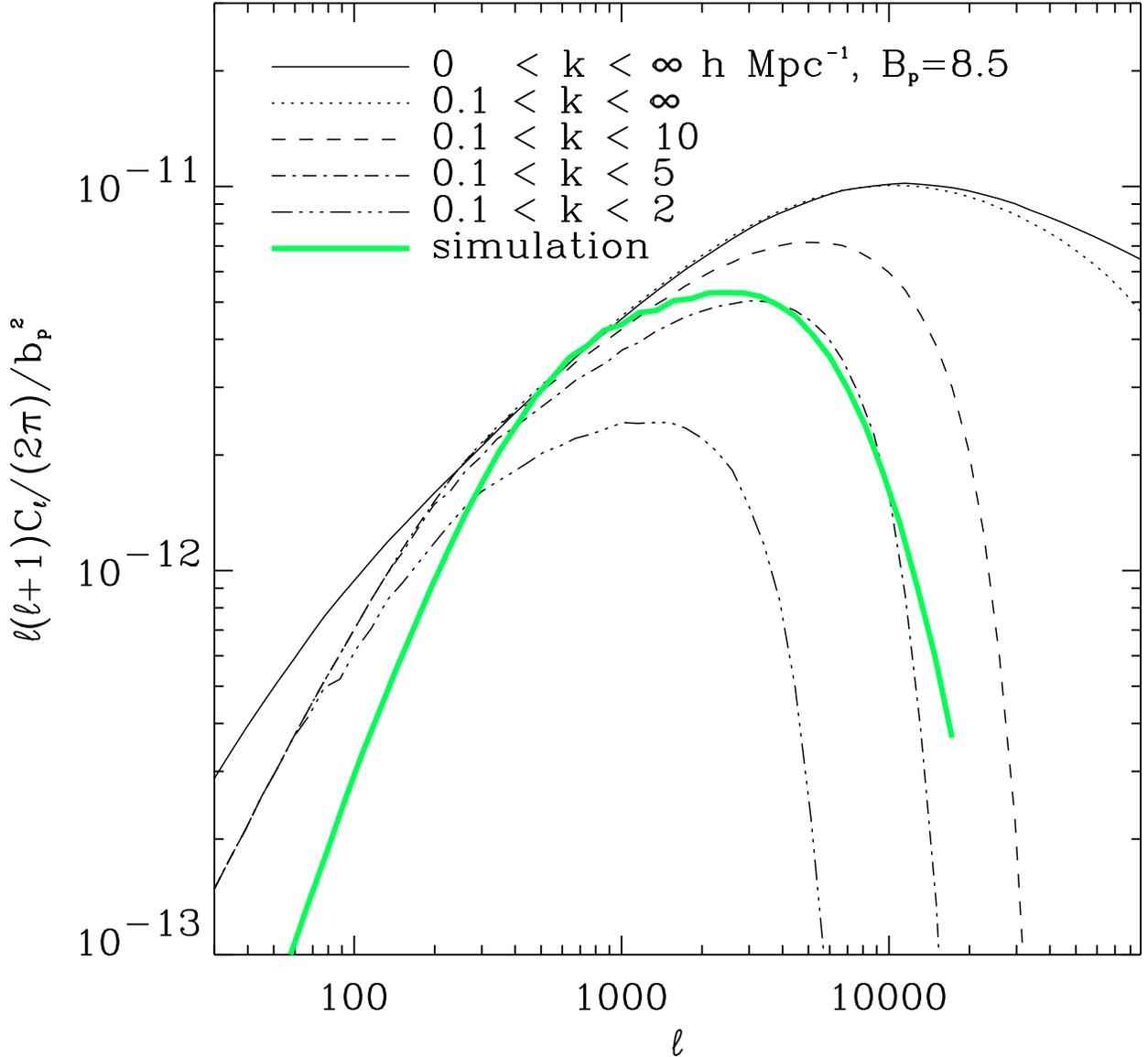}}
\caption{Effect of resolution and finite box size on the SZ power
spectrum. The $\Lambda$CDM power spectrum is shown for the constant
bias model (with $b_{p}=8.5$), for several restricted ranges of $k$
values for the pressure power spectrum $P_{p}(k,z)$. While the finite
box size ($0.1 < k < \infty \ h$ Mpc$^{-1}$) does not have much effect,
the finite resolution of the simulations ($0.1<k<2,5,10 \ h$ Mpc$^{-1}$)
reduce the full power spectrum ($0<k< \infty $) considerably outside
of the approximate range $200 < l < 2000$.}
\label{fig:cl_kran}
\end{figure}

The finite box size and resolution also suppress power in the pressure
power spectrum.  As we saw in \S\ref{resolution} and
Figure~\ref{fig:pk_rhodm}, the simulations lack power for $k \lesssim
0.1$ and $k \gtrsim 2 \ h$ Mpc$^{-1}$.  To test the impact of this
suppression, we consider the constant bias model described in
\S\ref{bias_model}. The total SZ power spectrum for this model is
shown in figure~\ref{fig:cl_kran} as the solid line, for the
$\Lambda$CDM case. This figure also shows the results of performing
the same calculation, but after suppressing power in several ranges of
$k$ values.  The finite box size (keeping only modes with $k > k_{\rm
min} = 0.1\ h$ Mpc$^{-1}$) reduces $C_{l}$ slightly for $l \lesssim
200$ and $l \gtrsim 20000$, and thus does not have a very large
effect.  On the other hand, the finite resolution ($k<k_{\rm
max}=2,5,10\ h$ Mpc$^{-1}$) reduces $C_{l}$ considerably for $l
\gtrsim 2000$. 

The above results can be interpreted as follows.
At a given $l$, the limited $k$-range corresponds to a limited $z$-range,
$l/k_{\rm max} < r(z) < l/k_{\rm min}$.
Let us take $k_{\rm min}=0.1$ and $k_{\rm max}=
2\ h\ {\rm Mpc^{-1}}$, as relevant for the simulations.
Then, $l=100$, $l=1000$ and $l=10000$ correspond to $0.02 \lesssim z \lesssim
0.4$, $0.2 \lesssim z < \infty$ and $5 \lesssim z < \infty$, respectively. 
Since most of contributions to $C_l$ come from $z<0.5$,
the finite $k_{\rm min}$ decreases $C_l$ only at low $l$, while 
$k_{\rm max}$ does so over the entire $l$ range.

We conclude that the limitations of the simulations precludes us
from predicting the SZ power spectrum outside of the $200 \lesssim
l \lesssim 2000$ range.  These limits can only be improved by using
larger simulations.  It is however worth noting that there could be
more SZ power around $l \sim 10000$. This might then be detectable
by future interferometric CMB measurements that have angular
resolutions around $1'$, intermediate in scale between
the satellite missions and the planned millimeter experiments
(ALMA, LMSA).

\subsection{Prospects for CMB Experiments}
\label{prospects}

\begin{figure}
\vspace*{3cm}
\centerline{\epsfig{file=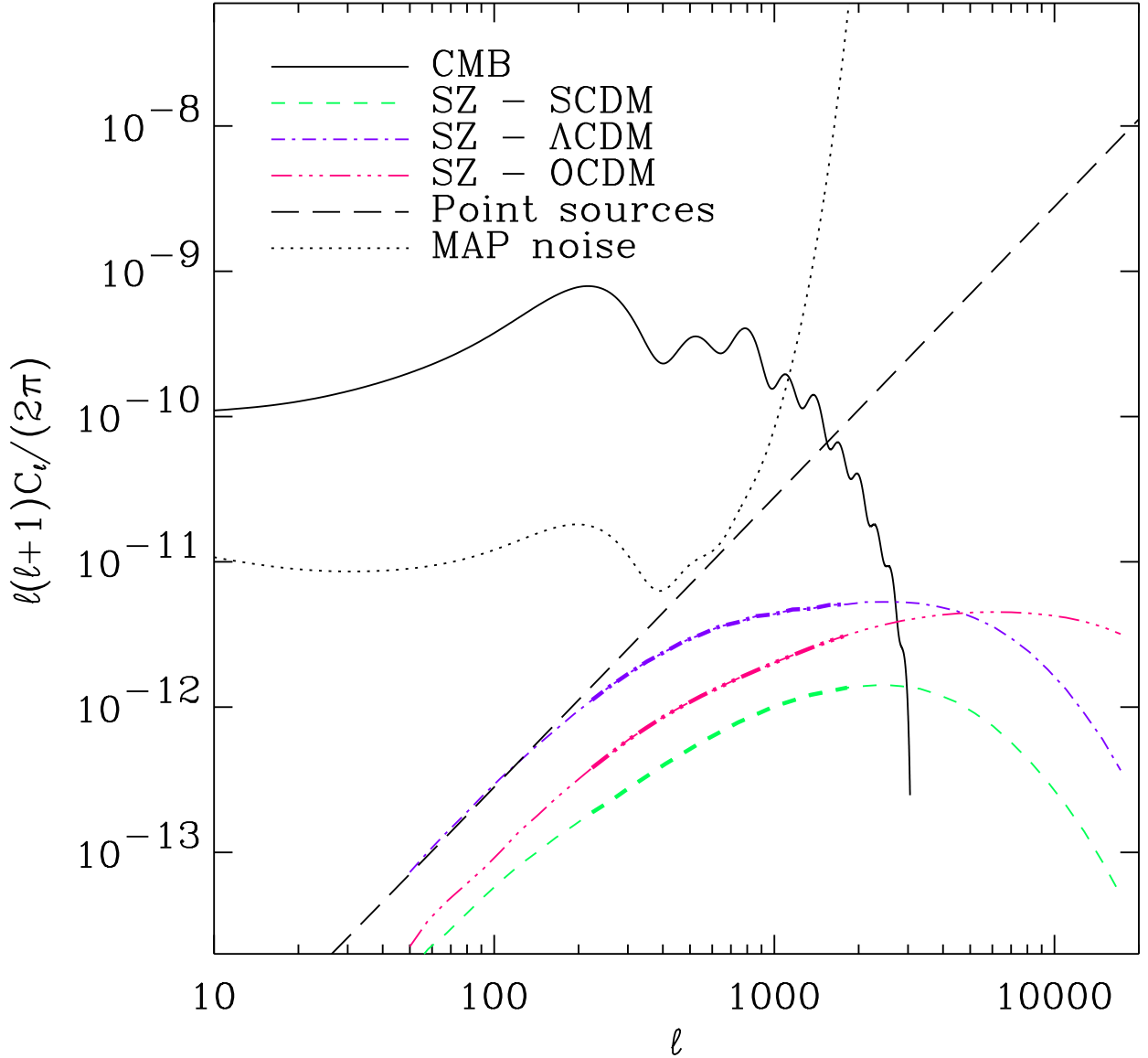}}
\caption{Power spectrum of the SZ effect for each model in the RJ
regime, as derived from the simulations. The approximate range of
confidence ($200 < l <  2000$) is highlighted by thicker lines.  The
power spectra outside of this range should be taken as lower limits.
For comparison, the primary CMB power spectrum is shown for the $\Lambda$CDM
model. The $1\sigma$ uncertainty for the 94GHz map channel is shown,
for a band average of $\Delta l = 10$. The power spectrum for the
residual discrete sources ($>2$Jy) for the 94 GHz MAP channel is also
shown.}
\label{fig:cl}
\end{figure}

The impact of secondary anisotropies on the upcoming MAP mission
\cite{ben95} were studied by Refregier et al.\cite{ref99b}. They showed
that discrete sources, gravitational lensing and the SZ effect were
the dominant extragalactic foregrounds for MAP. The dotted line on
Figure~\ref{fig:cl} shows the expected noise for measuring the primary
CMB power spectrum with the 94 GHz MAP channel, with a band average of
$\Delta l =10$. For all model considered, the SZ power spectrum is
well below the noise. The $rms$ $y$-parameter for the MAP 94 GHz beam
($13'$ FWHM) is listed in table~\ref{tab:results} for each model, from
both the simulations and the PS formalism. The resulting rms RJ
temperature fluctuations are of the order of a few $\mu$K, compared to a
nominal antenna noise of about $35\ \mu$K.  The SZ effect will therefore
not be a major limitation for estimating cosmological parameters with
MAP.

For comparison, the residual spectrum from undetected point sources
($S(94\ {\rm GHz})< 2$ Jy) expected using the model of Toffolatti et
al.\cite{tof98} is shown in figure~\ref{fig:cl} for
the 94 GHz channel. Point sources dominate over the SZ effect at $l
\gtrsim 300$, but are comparable below that.  Moreover, we have shown
in figure~\ref{fig:cl_z} and \ref{fig:cl_t}, 
that about 50\% of the SZ power spectrum at $l \simeq 500$ is 
produced at low redshifts ($z\lesssim 0.1$) and by
clusters of galaxies ($k_B T \gtrsim 5\ {\rm keV}$). 
This confirms the results of Refregier et al., 
who predicted that most of the SZ effect could be
removed by cross-correlating the CMB maps with existing X-ray cluster
catalogs (eg. XBACS \cite{ebe96}, BCS \cite{ebe97}).

Because of its limited spectral coverage, the MAP mission will not
permit a separation between the SZ effect and primordial
anisotropies. Apart from a handful of clusters which will appear as
point sources, it will therefore be difficult to detect the SZ
fluctuations directly with MAP. On the other hand, the future Planck
surveyor mission \cite{ber96} will cover both the positive and the
negative side of the SZ frequency spectrum, and will thus allow a
clear separation of the different foreground and background components
\cite{ber96,teg96,hob98,bou99}. Aghanim et al. \cite{agh97} have
established that, using such a separation, the SZ profiles of
individual clusters can be measured down to $y \simeq 3 \times
10^{-7}$. Moreover, Hobson et al. \cite{hob98} have estimated that the
SZ power spectrum could be measured for $50 \lesssim l \lesssim 1000$,
with a precision per multipole of about 70\%. Planck surveyor will
therefore provide a precise measurement of the total SZ power
spectrum. This would provide a direct, independent measurement of
$\Omega_{b}$ and of $\sigma_{8}$\cite{kom99}, and would thus help
breaking the degeneracies in the cosmological parameters estimated
from primordial anisotropies alone. Note that this measurement might
be also feasible, albeit with less precision, with upcoming balloon
experiments which also have broad spectral coverage.
\label{planck}

\subsection{The Missing Baryon Problem and Feedback}

The measured abundance of deuterium in low metallicity systems,
together with Big Bang Nucleosynthesis, predicts about twice as many
baryons than what is observed in galaxies, stars, clusters and neutral
gas \cite{cen99,fuk98}. These ``missing baryons'' are likely to be in
the form of the warm gas in groups in filaments \cite{pen99}. This
component is indeed difficult to observe directly since it is too cold
to be seen in the X-ray band, and too hot to produce any absorption
lines in the quasar spectra \cite{per98}.

The SZ effect on large scale could however provides a unique probe of
this warm gas. One can indeed imagine subtracting the detected
clusters from SZ maps, and measuring the power spectrum of the
residual SZ fluctuations, which are mainly produced by groups and
filaments.  For instance, if all clusters with $k_{B}T\lesssim 3$ keV were
removed from the SZ map, the SZ power spectrum would drop by a factor
of about 2 for $l \lesssim 2000$ (see figure~\ref{fig:cl_t}). For the
Planck Surveyor sensitivity quoted in \S\ref{planck}, this yields a
signal-to-noise ration per multipole of about 1. The amplitude, if not
the shape, of the residual SZ spectrum will thus be easily detected by
Planck, thus yielding constraints on the temperature and density of the
missing baryons.

In our simulations, we have only included gravitational forces.
However, feedback from star and AGN formation can also significantly
heat the IGM and thus affect the observed SZ effect. Valegeas \& Silk
\cite{val99} (see also reference therein), have studied the energy
injection produced by photo-ionization, supernovae, and AGN. In their
model, AGN are the most efficient, and can heat the IGM by as much as
$10^{6}$ K by a redshift of a few. This results in a mean
$y$-parameter of about $10^{-6}$, which is comparable to our value
derived from gravitational instability alone. Preheating by feedback
can thus increase the amplitude of the SZ effect by a factor of a few,
and thus be easily detected by Planck. Feedback can thus be directly
measured as an excess in the $y$-parameter or in the SZ power
spectrum, over the prediction from gravitational instability alone.
Energy injection has a large effect on the gas in groups and
filaments, comparatively to that in clusters. We may thus also detect
the effects of feedback through the relationship between the X-ray
temperature of groups (or their galaxy velocity dispersion) and their
SZ temperature. These measurements would then constrain the physics of
energy injection.

\section{Conclusions}
\label{conclusions}

We have studied the SZ effect using MMH simulations.  Our results for
the mean comptonization parameter is consistent with earlier work
using the Press-Schechter formalism and hydrodynamical simulations. It
is found to be lower than the current observational limit by about one
order of magnitude, for all considered cosmologies. The SZ power
spectrum is found to be comparable to the primary CMB power spectrum
at $l \sim 2000$. For the SCDM model, our SZ power spectrum is
approximately consistent with that derived by Persi et
al.\cite{per95}, after rescaling for the differing values of
$\sigma_{8}$.  We found that groups and filaments ($k_B T
\lesssim 5\ {\rm keV}$) contribute about 50\% of the SZ power
spectrum at $l=500$. On these scales, about 50\% of the SZ
power spectrum is produced at $z \lesssim 0.1$ and can
thus be removed using X-ray cluster catalogs. The SZ fluctuations are
well below the instrumental noise expected for the upcoming MAP mission,
and should therefore not be a limiting factor. The SZ power spectrum
should however be accurately measured by the future Planck
mission. Such a measurement will yield an independent measurement of
$\Omega_{b}$ and $\sigma_{8}$, and thus complement the measurements of
primary anisotropies\cite{kom99}.

We have compared our simulation results with predictions from the PS
formalism. The results from the two methods agree approximately, but
differ in the details.  The discrepancy could be due to the finite
resolution of the simulations, which limit the validity of our
predictions outside the $200 \lesssim l \lesssim 2000$ range. We also
find discrepancies with other numerical simulations. These issues can
only be settled with larger simulations, and by a detailed comparison
of different hydrodynamical codes. Such an effort is required for our
theoretical predictions to match the precision with which the SZ power
spectrum will be measured in the future.

A promising approach to measure the SZ effect on large scales is to
cross-correlate CMB maps with galaxy catalogs \cite{ref99a}. Most of
the SZ fluctuations on MAP's angular scales ($l < 1000$) are produced
at low redshifts and are thus correlated with tracers of the local
large scale structure.  Preliminary estimates indicate that such a
cross-correlation between MAP and the existing APM galaxy catalog
would yield a significant detection. Of course, even larger signals
are expected for the Planck Surveyor mission.  This would again
provide a probe of the gas distributed not only in clusters, but also
in the surrounding large-scale structure and therefore help solve the
missing baryon problem. Moreover, energy injection from star and AGN
formation can produce an SZ amplitude in excess of our predictions,
which only involve gravitational forces. The measurement of SZ
fluctuations or of a cross-correlation signal thus provides a measure
of feedback and can thus shed light on the process of galaxy
formation.

\acknowledgments 

We thank Uros Seljak and Juan Burwell for useful collaboration and
exchanges. We also thank Renyue Cen, Greg Bryan, Jerry Ostriker,
Arielle Phillips and Roman Juszkiewicz for useful discussions and
comparisons.  A.R. was supported in Princeton by the NASA MAP/MIDEX
program and the NASA ATP grant NAG5-7154, and in Cambridge by an EEC
TMR grant and a Wolfson College fellowship. D.N.S. is partially
supported by the MAP/MIDEX program.  E.K. acknowledges a fellowship
from the Japan Society for the Promotion of Science.  Computing
support from the National Center for Supercomputing applications is
acknowledged.  U.P. was supported in part by NSERC grant 72013704.
   

\end{multicols}

\newpage
\vspace*{7cm}
\begin{table}
\begin{tabular}{lrrrrrrrc}
Model & $\Omega_{m}$ & $\Omega_{\Lambda}$ & $\Omega_{b}$ &
$h$ & $\sigma_{8}$& $\Gamma$ & $N^{a}$ &
$L^{b}$ \\
\tableline 
SCDM         & 1.00 & 0.00 & 0.100 & 0.5 & 0.5 & 0.50 & $128^{3}$ & ~80 \\ 
$\Lambda$CDM & 0.37 & 0.63 & 0.049 & 0.7 & 0.8 & 0.26 & $128^{3}$ & 100 \\  
OCDM         & 0.37 & 0.00 & 0.049 & 0.7 & 0.8 & 0.26 & $128^{3}$ & ~80
\end{tabular}
\footnotesize{
$^{a}$ Number of curvilinear cells  \\
$^{b}$ Box Size ($h^{-1}$ Mpc)}
\caption{Simulation Parameters \label{tab:simulations}}
\end{table}

\vspace*{4cm}
\begin{table}
\begin{tabular}{l|cccc|ccc}
 & \multicolumn{4}{c|}{simulations} & \multicolumn{3}{c}{Press--Schechter}\\
Model & $\overline{T}_{\rho}^{a}$ (keV) &
$\overline{y} \times 10^{6}$ &
$\sigma_{y}(13') \times 10^{6}$ & $b_{p}^{a,b}$ & 
$\overline{T}_{\rho}^{a}$ (keV) & $\overline{y} \times 10^{6}$ & 
$\sigma_{y}(13') \times 10^{6}$ \\
\tableline SCDM & 0.19 & 0.86 & 0.33  & 5.30  & 0.27 & 0.86 & 0.58 \\ 
$\Lambda$CDM & 0.25 & 1.67 & 0.78 &  9.51 & 0.39 & 2.11 & 1.21 \\
OCDM & 0.19 & 2.62 & 0.45 & 4.54 & 0.42  & 3.23  & 1.57
\end{tabular}
\footnotesize{
$^{a}$ at $z=0$\\
$^{b}$ for $k=0.5 \ h$ Mpc$^{-1}$}
\caption{Results \label{tab:results}}
\end{table}

\end{document}